\documentclass[12pt]{iopart}
\usepackage{graphicx}

\begin{document}

\title[How strange are compact star interiors ?]
{How strange are compact star interiors ?}

\author{D.~Blaschke$^{1,2}$, T.~Kl\"ahn$^1$, R.~{\L}astowiecki$^1$ and 
F.~Sandin$^{3}$}

\address{$^{1}$Institute for Theoretical Physics, 
University of Wroclaw, 50-204 Wroclaw, Poland}
\address{$^{2}$Bogoliubov Laboratory for Theoretical Physics,
JINR Dubna, 141980 Dubna, Russia}
\address{$^{3}$EISLAB, Lule{\aa} University of Technology, 97187 Lule{\aa}, 
Sweden}
\ead{blaschke@ift.uni.wroc.pl}

\begin{abstract}
We discuss a Nambu--Jona-Lasinio (NJL) type quantum field theoretical approach 
to the quark matter equation of state with color superconductivity and 
construct hybrid star models on this basis.
It has recently been demonstrated that with increasing 
baryon density, the different quark flavors may occur sequentially, starting 
with down-quarks only, before the second light quark flavor and at highest 
densities also the strange quark flavor appears. 
We find that color superconducting phases are favorable over 
non-superconducting ones which entails consequences for thermodynamic and 
transport properties of hybrid star matter.
In particular, for NJL-type models no strange quark matter phases
can occur in compact star interiors due to mechanical instability against
gravitational collapse, unless a sufficiently strong flavor mixing as provided 
by the Kobayashi-Maskawa-'t Hooft determinant interaction is present in the 
model.
We discuss observational data on mass-radius relationships 
of compact stars which can put constraints on the properties of dense matter 
equation of state.
\end{abstract}

\maketitle

\section{Introduction}

The question of strangeness in compact star interiors is a multifaceted one 
\cite{Weber:2004kj} and has been discussed controversially ever since the 
first discussion of hyperons  \cite{Ambartsumyan:1960} and kaon condensates 
\cite{Brown:1988ik,Brown:1992ib,Glendenning:1998zx} 
for the degenerate matter phases of neutron star cores.
Present-day state-of-the-art calculations of hyperonic matter within the 
Brueckner-Bethe-Goldstone scheme \cite{Baldo:1999rq}
reveal the problem that due to the softening
of the high-density equation of state (EoS) the maximum mass of hyperonic 
stars is below the measured mass of binary radiopulsars 
$M_{\rm BRP}=1.35 \pm 0.04~M_\odot$ \cite{Thorsett:1998uc}.   
As a possible solution to this problem a sufficiently early onset of quark 
deconfinement in neutron star matter at baryon densities 
$n= 0.3 - 0.5$ fm$^{-3}$ has been suggested \cite{Baldo:2003vx}.
In the present contribution we will not touch the question of absolutely 
stable strange quark matter \cite{Witten:1984rs} in compact stars but rather 
focus the discussion on models for 
hybrid stars with deconfined quark matter cores and recent observational 
constraints for their masses ($M$) and radii ($R$), see 
\cite{Lattimer:2006xb,Klahn:2006ir}.

The status of the theoretical approach to the neutron star matter equation of
state  is very different from that for the high-temperature case at low or 
vanishing net baryon densities, where {\it ab initio} lattice QCD simulations 
provide EoS with almost physical quark masses systematically approaching the 
continuum limit \cite{Bazavov:2009zn}.
This guidance is absent at zero temperature and high baryon number densities,
where a variety of models exists on different levels of the microphysical 
description which make different predictions for the state of matter.  
A common feature is that the transition from hadronic to quark matter cannot 
yet be described on a unique footing where hadrons would appear as bound states
(clusters) of quarks and their possible dissociation at high densities 
as a kind of Mott transition \cite{Mott:1968} like in nonideal plasmas 
\cite{Redmer:1997} or in nuclear matter \cite{Ropke:1982,Typel:2009sy}. 
Early nonrelativistic potential model approaches 
\cite{Horowitz:1985tx,Ropke:1986qs} are insufficient since they 
cannot accomodate the chiral symmetry restoration transition in a proper way.
Therefore, at present the discussion is restricted to so-called two-phase 
aproaches where the hadronic and the quark matter EoS are modeled separately
followed by a subsequent phase transition construction to obtain a hybrid EoS.

Widely employed for a description of quark matter in compact stars are 
thermodynamical bag models of three-flavor quark matter with 
eventually even density-dependent bag pressure $B(n)$, as in Ref. 
\cite{Baldo:2003vx}. 
A qualitatively new feature of the phase structure appears
in chiral quark models of the Nambu--Jona-Lasinio type which describe the 
dynamical chiral symmetry breaking of the QCD vacuum and its partial 
restoration in hot and dense matter, see Ref. \cite{Buballa:2003qv} for a 
review. In contrast to bag models, in these approaches at low temperatures 
the light and strange quark degrees of freedom may appear sequentially with 
increasing density \cite{Gocke:2001ri,Blaschke:2008vh,Blaschke:2008br}, 
so that strangeness may even not 
appear in the quark matter cores of hybrid stars, before their maximum mass
is reached. 
Once chiral symmetry is restored, a rich spectrum of color superconducting 
quark matter phases may be realized in dense quark matter, depending on it's 
flavor composition and isospin asymmetry  \cite{Alford:2007xm} with 
far-reaching consequences for hybrid star phenomenology, e.g., $M-R$ 
relationships and cooling behavior. 

We will consider here a color superconducting three-flavor NJL model with 
selfconsistently determined density dependences of quark masses and scalar 
diquark gaps, developed in Refs.  \cite{Ruester:2005jc} and 
\cite{Blaschke:2005uj} which differ in the fact that the former includes the 
flavor-mixing Kobayashi-Maskawa-'t Hooft (KMT) determinant interaction 
\cite{Kobayashi:1970ji,'tHooft:1976up}
while the latter does not.   
Only recently it became clear \cite{Agrawal:2010er} that this flavor mixing
is crucial for the possible stability of strange quark matter phases in 
hybrid stars.
 
\section{Role of the Kobayashi-Maskawa-'t Hooft (KMT) interaction}

Our description of quark matter is based on the grand canonical thermodynamic 
potential 
~\cite{Buballa:2003qv,Ruester:2005jc,Blaschke:2005uj,Abuki:2005ms,Klahn:2006iw}
\begin{eqnarray}
	\Omega(T,\{\mu\}) &=& \frac{\phi^2_u+\phi^2_d+\phi^2_s}{8 G_S}
	+\frac{K\phi_u\phi_d\phi_s}{16G^3_S} 
	-\frac{\omega^2_u+\omega^2_d+\omega^2_s}{8 G_V}
	+\frac{\Delta^2_{ud}+\Delta^2_{us}+\Delta^2_{ds}}{4 G_D} 
\nonumber \\
	&-&\int\frac{d^3p}{(2\pi)^3}\sum_{n=1}^{18}
	\left[E_n+2T\ln\left(1+e^{-E_n/T}\right)\right] 
	+ \Omega_{lep} - \Omega_0~,
\label{potential}
\end{eqnarray}
where $E_n=E_n(p,\,\mu;\,\mu_Q,\mu_3,\mu_8,\,\phi_u,\phi_d,\phi_s,\,
\omega_u,\omega_d,\omega_s,\,\Delta_{ud},\Delta_{us},\Delta_{ds})$ 
are the quasiparticle dispersion relations, 
obtained by numerical diagonalization of the quark propagator matrix in 
color-, flavor-, Dirac- and Nambu-Gorkov spaces.
The values of the meanfields (order parameters) are obtained from a 
minimization of $\Omega(T,\{\mu\})$, which is equivalent to the selfconsistent 
solution of the set of corresponding gap equations. 
$\Omega_{lep}$ is the contribution from leptons ({\it e.g.}, electrons, muons 
and the corresponding neutrino flavors) and the subtraction of $\Omega_0$ 
garantees that in the vacuum $\Omega(0,\,0)=0$.
Eq. (\ref{potential}) is obtained in the meanfield approximation to the path 
integral representation of the partition function for the three-flavor NJL 
model Lagrangian with the interaction channels
\begin{eqnarray}
	{\cal L}_{\bar{q}q} &=& G_S \sum_{a=0}^8 \Big[({\bar q}\tau_a q)^2
	+ ({\bar q} i\gamma_5\tau_a q)^2\Big] 
	+ G_V({\bar q} i\gamma_0 q)^2
\nonumber \\
	&-&K\left[{\rm det}_f(\bar{q}(1+\gamma_5)q)
	+{\rm det}_f(\bar{q}(1-\gamma_5)q)\right],
\label{qbarqlag} \\
	{\cal L}_{qq} &=& G_D\!\!\!\!\!\!\sum_{a,b = 2,5,7}\!\!\!\!\!\! 
	(\bar{q} i\gamma_5 \tau_a \lambda_b C\bar{q}^T)
	(q^T C i\gamma_5\tau_a\lambda_b\,q),
\label{qqlag}
\end{eqnarray}
where $\tau_a$ and $\lambda_b$ are the antisymmetric Gell-Mann matrices acting 
in flavor and color space, respectively. 
The scalar ($G_S$), diquark ($G_D$), vector ($G_V$) and KMT ($K$) couplings are
to be determined by hadron phenomenology, see \cite{Rehberg:1995kh}. 

\begin{figure} [!ht] 
\includegraphics[width=0.5\textwidth]{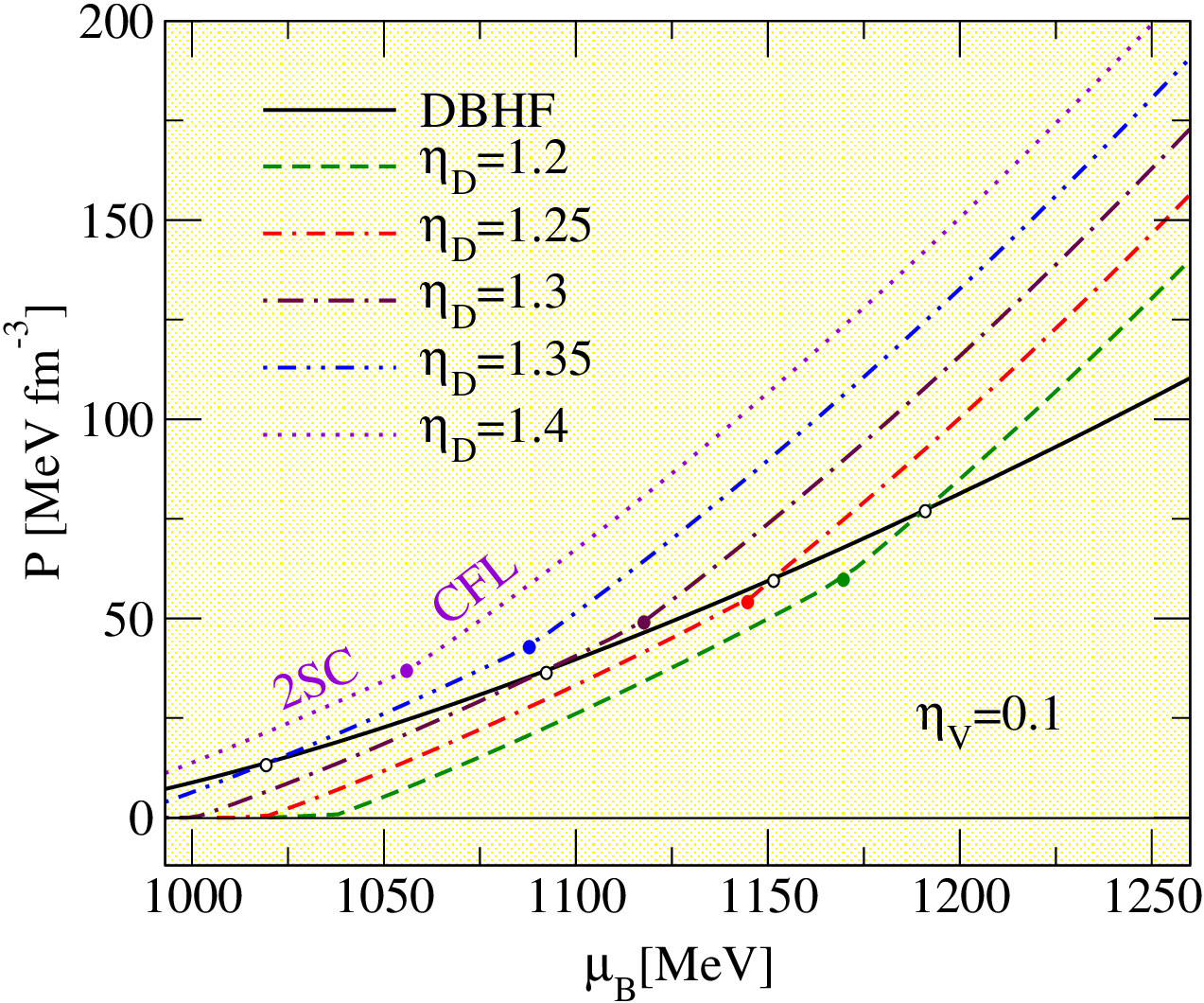}
\includegraphics[width=0.5\textwidth]{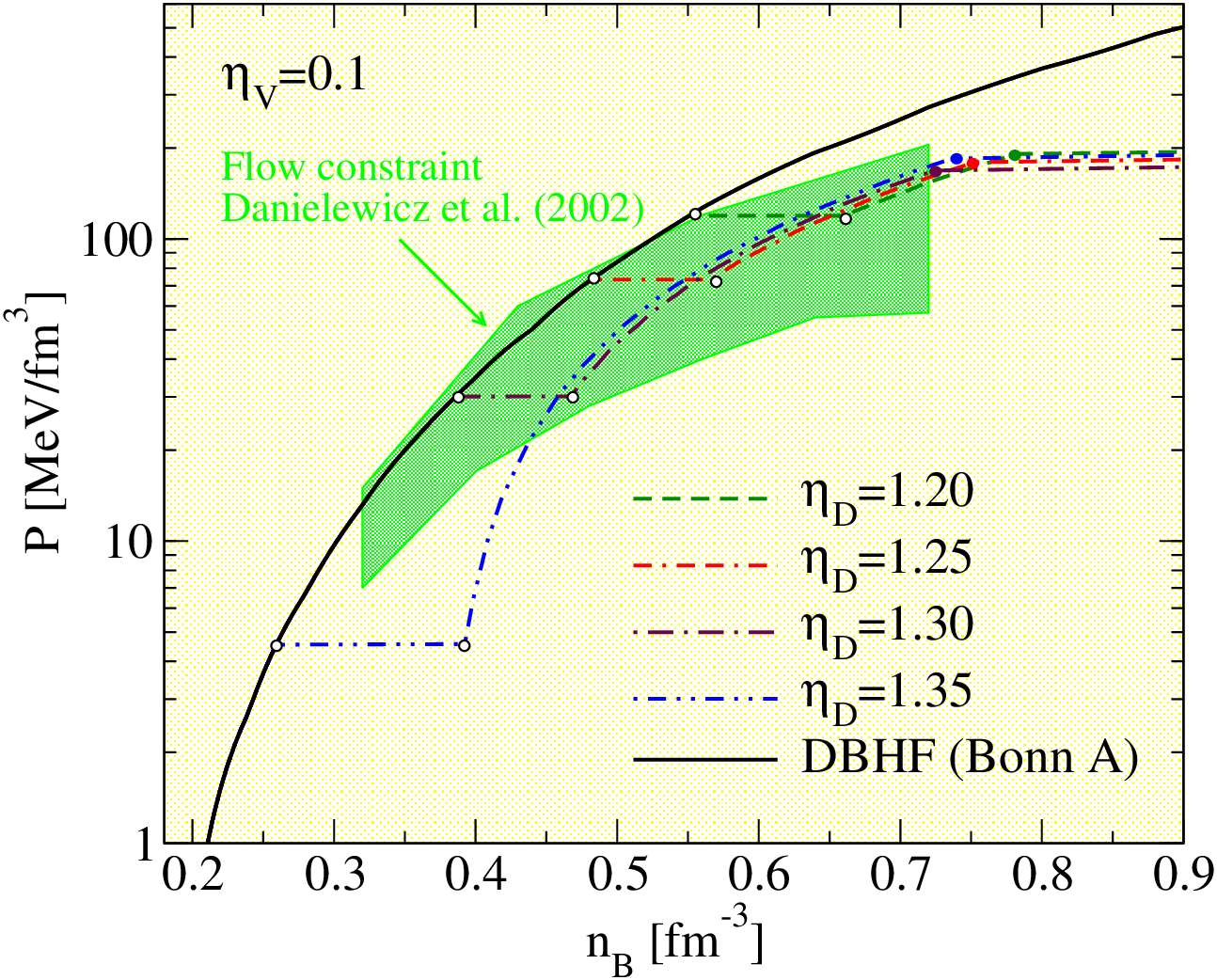}
\caption{Left: Equations of state for neutron star matter in beta-equilibrium. 
The hadronic phase is described by the DBHF EoS (solid line) and the 
quark matter EoS correspond to the NJL model (\ref{potential}) for different 
choices of the diquark coupling strength $\eta_D=G_D/G_S$ for given vector 
coupling $\eta_V=G_V/G_S=0.1$ and KMT coupling $K$. 
Phase transitions from two-flavor (2SC) to three-flavor 
(CFL) quark matter are marked by full dots whereas 
phase transitions from hadronic to quark matter are indicated by open dots. 
Right: Hybrid equations of state for symmetric matter compared to the
flow constraint \cite{Danielewicz:2002pu} from heavy-ion collisions.
Except $\eta_D=1.35$, for which the onset of the 2SC phase has a too low 
critical density, the parametriztions fulfil the flow constraint, see also 
\cite{Lattimer:2006xb,Klahn:2006ir,Klahn:2006iw}.
The 2SC-CFL transition lies outside the tested density region.} 
\label{f:EoS} 
\end{figure} 

As pointed out in \cite{Buballa:2003qv}, due to the mixing of the light and 
strange flavor sectors by the KMT interaction, the difference in the critical 
chemical potentials for the chiral phase transitions in these sectors 
(which coincide with the onset of 2SC and CFL phases, respectively) gets 
diminished. 
This entails that the phase transition between hadronic matter 
(described by a realistic nuclear EoS, e.g., the DBHF one, see 
\cite{Klahn:2006iw}) and superconducting quark matter may eventually proceed 
directly into the CFL phase, provided the diquark coupling is sufficiently 
strong, see the left panel of Fig.~\ref{f:EoS}. 
For consistency with nuclear matter phenomenology, one may check that the 
corresponding isospin symmetric EoS 
would not predict a too low transition density and is not in contradiction with
the flow constraint from heavy-ion collisions \cite{Danielewicz:2002pu}, see
the right panel of Fig.~\ref{f:EoS}.  

\begin{figure} [!ht] 
\includegraphics[width=\textwidth]{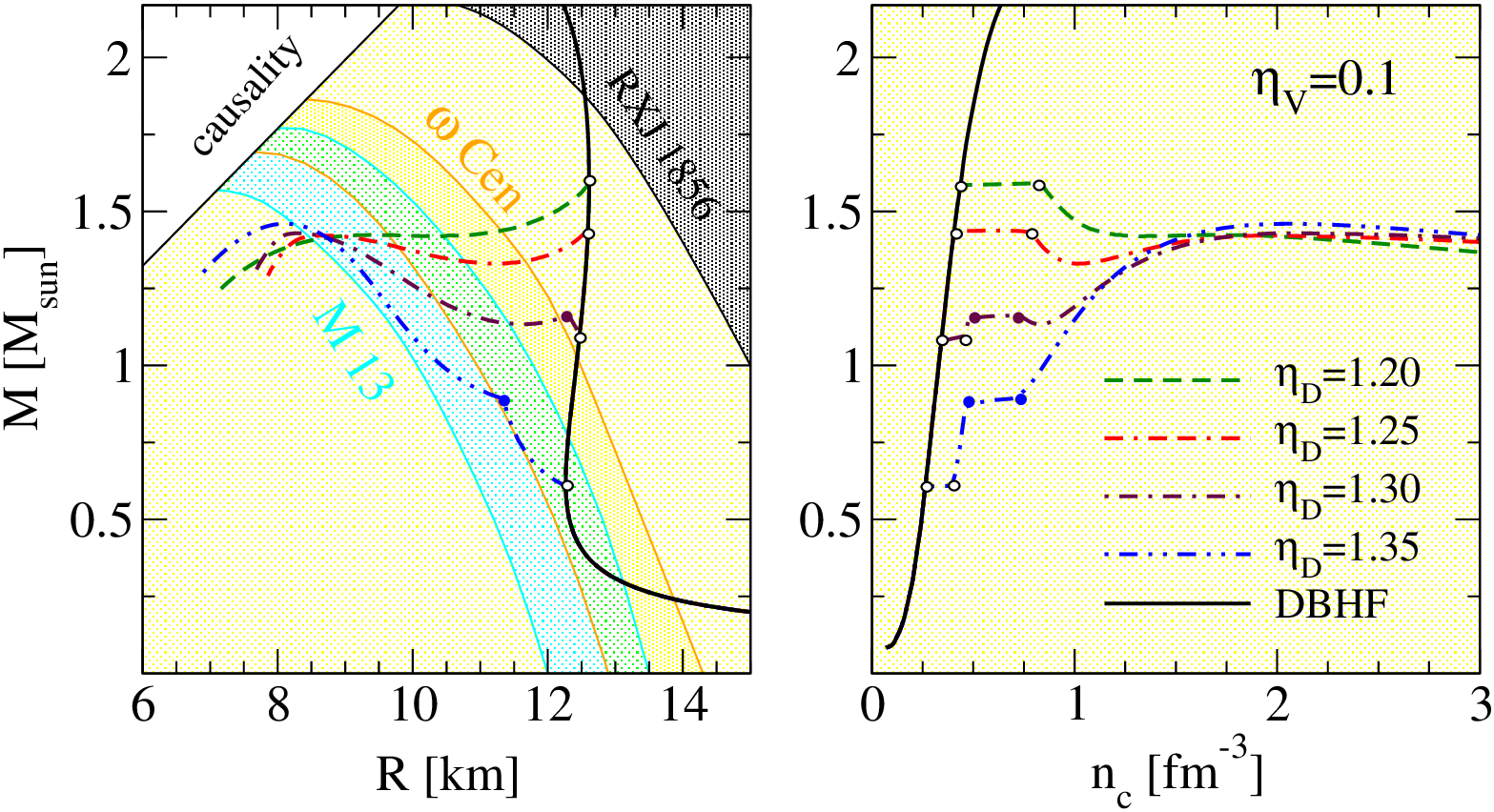}
\caption{Compact star sequences for different values of the diquark coupling
strength $\eta_D$. For $\eta_D=1.2$ no stable configurations with quark matter 
exist while for the strong coupling $\eta_D=1.35$ there is an early onset of
2SC quark core configurations at $M=0.6~M_\odot$ followed by a stable branch 
of hybrid stars CFL quark matter cores in the mass range $0.9<M/M_\odot<1.46$.
For comparison recent mass radius constraints are shown coming from
RX J1856 \cite{Trumper:2003we,Klahn:2006ir},  M13 \cite{Gendre:2003pw} and 
$\omega$ Cen \cite{Gendre:2002rx}, see also \cite{Lattimer:2006xb}.
}
\label{f:M-R} 
\end{figure} 

In Fig.~\ref{f:M-R} we show the $M-R$ and $M-n_c$ relationships for nonrotating
compact star sequences obtained as solutions of the Tolman-Oppenheimer-Volkoff 
equations for the hybrid EoS shown in the left panel of Fig.~\ref{f:EoS} where
the diquark coupling $\eta_D=G_D/G_S$ is a free parameter and the vector 
coupling is set to $\eta_V=G_V/G_S=0.1$. For sufficiently large $\eta_D>1.2$ 
there is a stable branch of hybrid stars with color superconducting quark 
matter core. These solutions are more compact and have a higher central density
$n_c$ than the purely hadronic stars with the same mass. In the parameter range
$1.25<\eta_D<1.3$, mass twin solutions are possible, for which the more 
compact configuration has a CFL quark matter core. 
Such solutions have not been possible without the KMT interaction term, when
the occurrence of CFL quark matter marks the onset of gravitational instability
and thus sets the limit for the maximum mass of the hybrid star 
\cite{Klahn:2006iw,Blaschke:2007tt}.     
The question arises whether other flavor-mixing interactions have to be 
invoked. It seems necessary in a next step to include channels which appear 
after Fierz transformation of the KMT interaction and couple 
chiral condensates with diquark condensates  \cite{Steiner:2005jm}.
This may lead to a new critical point in the QCD phase diagram 
\cite{Hatsuda:2006ps} and, depending on the sign of the coupling, to a further
reduction of the strange chiral condensate which enforces the flavor mixing
effect studied here.  

The observational constraints for masses and radii are not yet settled. There
is a lower limit for the $M-R$ relation from RXJ 1856.5-3754
\cite{Trumper:2003we} which requires either a large radius $R>14$ km for a 
star with $M=M_{\rm BRP}$ or a mass larger than $\sim 2~M_\odot$ for a star 
with 12 km radius. 
Both is difficult to accomodate with the present EoS.
$M-R$ relations for the quiescent binary neutron stars in globular clusters
M13 \cite{Gendre:2003pw} and $\omega $ Cen  \cite{Gendre:2002rx} point to 
rather compact stars as described by those sequences obtained here.
The presented microscopic EoS features an intermediate
softening of the EoS due to diquark condensation in the density range
$\sim 0.4 - 0.8$ fm$^{-3}$ and predicts two stable branches in
an overlapping range of masses (twins) as a robust feature of compact star 
$M-R$ relationships.
The hypothesis of the existence of a ``third family'' of supercompact stars
is testable in future observational campaigns
devoted to determine the masses \cite{Freire:2009dr} and $M-R$ relationship 
for compact stars \cite{Ozel:2009da} with high precision and thus to constrain
the dense matter EoS. 
 
\section{Conclusions}

The effect of the flavor-mixing KMT determinant interaction on the sequential 
occurrence of superconducting two- and three-flavor quark matter phase 
transitions in the EoS for cold dense matter has been studied in an NJL-type
model. 
It is found that at sufficiently strong diquark coupling the deconfinement 
phase transition in neutron star matter leads directly from the hadronic phase 
to the CFL phase and occurs at low enough density to entail the formation 
of stable branch of CFL core hybrid stars in
the $M-R$ diagram of compact stars.
This branch may either continuously join that of normal neutron stars or 
eventually form a ``third family'' branch, separated from that of the neutron 
stars by a sequence of unstable configurations.
The possibility of such a characteristic feature of mass twins
in the $M-R$ diagram suggested by advanced microscopic QCD motivated hybrid 
EoS with superconducting dense quark matter phases should be kept in the focus 
of observational programs to deduce the cold dense EoS exploiting measured 
$M-R$ relationships for compact stars.

\section*{Acknowledgments}
DB and RL acknowledge the hospitality of the Yukawa Institute for Theoretical 
Physics Kyoto, partial support by the Yukawa International Program for
Quark-Hadron Sciences and discussions during the Workshop program 
``New Frontiers in QCD'' where this work was completed. 
This work has been supported in part by the Polish Ministry of Science and Higher 
Education  (MNiSW) under grant No. N N 202 2318 37 (DB, TK, RL), by 
the Russian Fund for Basic research (RFBR) under grant No. 08-02-01003-a (DB),
by FNRS, the Belgian fund for scientific research (FS) and by CompStar, a 
research networking programme of the European Science Foundation.

\end{document}